\documentclass[12pt]{article}
\usepackage[utf8]{inputenc}
\usepackage[small,bf]{caption}
\usepackage{fullpage,graphicx,psfrag,url,ar,rotating,url,hyperref}
\usepackage{amsmath,amssymb,amsthm,enumitem,subcaption,appendix}
\usepackage{makecell}

\usepackage[final]{changes}
\definechangesauthor[name=Reviewer1, color=blue]{R1}
\definechangesauthor[name=Reviewer2, color=blue]{R2}

\title{Distributed and Scalable Optimization for Robust Proton Treatment Planning}
\author{Anqi Fu \and Vicki T. Taasti \and Masoud Zarepisheh}
\date{\today}

\begin{document}
	
\maketitle

\begin{abstract}
\mbox{}\par\vspace{-\baselineskip}

\paragraph{Purpose:} The importance of robust proton treatment planning to mitigate the impact of uncertainty is well understood. However, its computational cost grows with the number of uncertainty scenarios, prolonging the treatment planning process. We developed a fast and scalable distributed optimization platform that parallelizes this computation over the scenarios.

\paragraph{Methods:} We modeled the robust proton treatment planning problem as a weighted least-squares problem. To solve it, we employed an optimization technique called the Alternating Direction Method of Multipliers with Barzilai-Borwein step size (ADMM-BB). We reformulated the problem in such a way as to split the main problem into smaller subproblems, one for each proton therapy uncertainty scenario. The subproblems can be solved in parallel, allowing the computational load to be distributed across multiple processors (e.g., CPU threads/cores). We evaluated ADMM-BB on four head-and-neck proton therapy patients, each with $13$ scenarios accounting for $3$ mm setup and $3.5\%$ range uncertainties. We then compared the performance of ADMM-BB with projected gradient descent (PGD) applied to the same problem.

\paragraph{Results:} For each patient, ADMM-BB generated a robust proton treatment plan that satisfied all clinical criteria with comparable or better dosimetric quality than the plan generated by PGD. However, ADMM-BB's total runtime averaged about $6$ to $7$ times faster. This speedup increased with the number of scenarios.

\paragraph{Conclusion:} ADMM-BB is a powerful distributed optimization method that leverages parallel processing platforms, such as multi-core CPUs, GPUs, and cloud servers, to accelerate the computationally intensive work of robust proton treatment planning. This results in 1) a shorter treatment planning process and 2) the ability to consider more uncertainty scenarios, which improves plan quality.
\end{abstract}

\section{Introduction}

Proton treatment planning has been an active topic of research over the last decade. The sharp dose fall-off of protons, which makes proton therapy an appealing treatment modality, also renders it vulnerable to errors and uncertainties during treatment planning. Consequently, proton plans are usually developed using robust optimization \cite{Unkelbach:2018}. The dose distribution for each potential uncertainty scenario (e.g., due to setup or range estimation errors) is computed, then the plan is optimized to fulfill clinical objectives even in these scenarios.

To produce the most robust plan, we would like to consider as many scenarios as possible, covering all potential sources of uncertainty. This presents a number of computational challenges. First, the more scenarios we include in the optimization problem, the longer it will take to solve the problem. This could prolong the treatment planning process and limit the number of parameter adjustments (e.g., objective weights, penalties) that we make, which may result in a suboptimal plan. Second, the memory required to handle all the desired scenarios could exceed the resources available to us, especially on a single computing platform. It is thus of great clinical importance to develop a fast, distributed, and scalable method for robust proton treatment planning.

Given the large size of robust optimization problems in proton therapy, first-order methods like gradient descent have been the \textit{de-facto} solution technique. In particular, Projected Gradient Descent (PGD) enjoys widespread popularity, as it can handle upper and lower bounds on spot intensities. However, while gradient descent is widely used, it possesses several weaknesses that make it undesirable for solving complex, data-intensive optimization problems. First, it requires the calculation of the gradient, which may be mathematically cumbersome or intractable. Second, its convergence depends heavily on the step size (i.e., the distance moved in the direction of the gradient), which is typically chosen via a line search. This line search is computationally demanding and scales poorly with the size of the problem. Moreover, the very serial format of the line search makes gradient descent unamenable to parallelization. Alternative methods exist for step size selection \cite{DuchiHazan:2011,KingmaBa:2014,TanMa:2016,QiaoLew:2016}, but in general, optimization practitioners turn to the class of distributed algorithms to solve very large problems.

\added[id=R1]{In distributed optimization, multiple agents (e.g., CPUs) collaborate to solve an optimization problem. A typical algorithm will decompose the problem into parts and distribute each part to an agent, which carries out its computation using local information. The agents then combine their results to produce a solution. Due to the use of many agents, these algorithms scale remarkably well: their computational demands increase very modestly with the size and complexity of the problem. Consequently, distributed optimization has been an active topic of research for decades \cite{BertsekasTsitsiklis:1989,Sayed:2014,NedicLiu:2018,YangJohansson:2019}, and distributed algorithms have been applied in a variety of fields \cite{RaffardBoyd:2004,RabbatNowak:2004,WangWu:2017,FuBoyd:2022}.}

The Alternating Direction Method of Multipliers (ADMM) is a distributed optimization algorithm dating back to the 1970's \cite{GabayMercier:1976,BoydParikh:2011,FangHe:2015}. It has seen renewed interest over the last few years thanks to its success in solving large-scale optimization problems that arise in data science and machine learning \cite{HuangHu:2019,GuFan:2018,RamdasTibshirani:2016,DharYi:2015,SawatzkyXu:2014}. The key to ADMM's success is its ability to split a problem into smaller subproblems, which can be solved independently from each other, making it ideal for parallel computation. A clever split can also yield mathematically simple subproblems that permit a closed-form solution. Recently, \cite{ZarepishehXing:2018} introduced a new variant of ADMM with improved convergence properties and evaluated its performance on fluence map optimization problems in Intensity Modulated Radiation Therapy (IMRT). They showed that ADMM outperforms other optimization techniques, including an active-set method (FNNLS), a gradient-based method (SBB), and an interior-point solver (CPLEX). This demonstrates the potential of ADMM to efficiently handle large treatment planning problems.

Our paper extends the application of ADMM to robust proton treatment plan optimization. In particular, we exploit the special structure of the robust optimization problem, which enables us to reformulate the problem and split it into smaller subproblems, each corresponding to a separate uncertainty scenario. A major difference between our study and \cite{ZarepishehXing:2018} is that we implement a {\em fully parallelized} version of ADMM that distributes the workload required to solve these subproblems across multiple CPU threads/cores. This allows us to achieve a greater and more consistent speed advantage over PGD. More importantly, we show that our ADMM algorithm scales well with the number of scenarios, making it possible to include many different sources of uncertainty in the treatment planning process.

\section{Methods and materials}
\label{sec:methods}

\subsection{Problem formulation and gradient descent}
\label{sec:problem}

Suppose we have $N$ scenarios, including the nominal scenario, represented by dose-influence matrices $A_s \in \mathbf{R}_+^{m \times n}$ for $s = 1,\ldots,N$. Here $m$ is the number of voxels and $n$ is the number of beamlets. Let $a_{s,i} \in \mathbf{R}_+^n$ be row $i$ of $A_s$, corresponding to voxel $i$. The prescribed dose to each voxel is given in the form of a vector $p \in \mathbf{R}_+^m$, where $p_i$ is equal to the prescription (i.e., a constant scalar $D_{\text{pres}} > 0$) for target voxels and zero for non-target voxels. Our objective is to find spot intensities $x \in \mathbf{R}^n$ that minimize the deviation between the actual and prescribed dose across all scenarios. 

Formally, let us define the {\em scenario-specific objective function} to be
\begin{equation}
	\label{eq:obj_scenario}
	f_s(x) := \sum_{i=1}^m w_i\|a_{s,i}^Tx - p_i\|_2^2,
\end{equation}
where $w_i \geq 0$ is the relative importance weight on voxel $i$. While in principle, different voxels within a structure can admit different weights, in our model, we assign the same weight to all voxels in a structure and only allow weights to vary across structures. Our goal is to solve the optimization problem
\begin{equation}
	\label{prob:nnls_full}
	\begin{array}{ll}
		\mbox{minimize} & f(x) := \sum_{s=1}^N f_s(x) \\
		\mbox{subject to} & x \geq 0
	\end{array}
\end{equation}
with variable $x \in \mathbf{R}^n$. \added[id=R1]{Here we have assumed all scenarios occur with equal probability. It is straightforward to incorporate differing probabilities by replacing $w_i$ with scenario-specific weights $\tilde w_{s,i} := c_sw_i$, where $c_s \in [0,1]$ is the probability of scenario $s$.}

PGD starts from an initial estimate of the spots $x^{(0)} \in \mathbf{R}_+^n$ and iteratively computes
\begin{align*}
	\nabla f(x^{(k)}) &= \sum_{s=1}^N \sum_{i=1}^m 2w_ia_{s,i}(a_{s,i}^Tx^{(k)} - p_i), \\
	x^{(k+1)} &= \max(x^{(k)} - \rho^{(k)} \nabla f(x^{(k)}), 0), 
\end{align*}
where $\rho^{(k)} > 0$ is the step size in iteration $k$, selected using a line search technique such as \textit{Armijo} \cite[Section 3.1]{NocedalWright:2006} to ensure improvement in the objective value. \textit{Armijo} line search starts from an initial step size $\rho^{(k)} = \alpha$ and shrinks the step size by a factor of $\gamma$ until it produces $x^{(k+1)}$ that satisfies
\begin{equation}
	\label{eq:armijo_cond}
	f(x^{(k+1)}) \geq f(x^{(k)}) + \beta \nabla f(x^{(k)})^T(x^{(k+1)} - x^{(k)}).
\end{equation}
Here $\alpha > 0, \beta \in (0,1)$, and $\gamma \in (0,1)$ are fixed parameters. PGD proceeds until some stopping criterion is reached, typically when the change in the objective value, $|f(x^{(k)}) - f(x^{(k+1)})|$, falls below a user-defined cutoff.

The gradient of each scenario's objective function, 
\begin{equation}
	\label{eq:grad_scenario}
	\nabla f_s(x^{(k)}) = \sum_{i=1}^m 2w_ia_{s,i}(a_{s,i}^Tx^{(k)} - p_i),
\end{equation}
can be computed in parallel, i.e., simultaneously on different processing units (CPUs, GPUs, etc), and summed up to obtain the gradient of the full objective, $\nabla f(x^{(k)})$. However, the \textit{Armijo} condition \ref{eq:armijo_cond} must be checked and updated serially, creating a bottleneck in any parallel implementation of PGD.

\subsection{Distributed alternating direction method of multipliers}
\label{sec:admm}

To apply ADMM to Problem (\ref{prob:nnls_full}), we must first reformulate it so the objective function is separable. The current objective $f(x)$ consists of a sum of scenario-specific objectives $f_s(x)$, which share the same variable $x$. We will replace $x$ with new variables $x_1,\ldots,x_N \in \mathbf{R}^n$, representing the scenario-specific spots, and a so-called consensus variable $z \in \mathbf{R}^n$. We then introduce a consensus constraint $x_s = z$ for $s = 1,\ldots,N$ to ensure the spot intensities are equal. This results in the mathematically equivalent formulation
\begin{equation}
	\label{prob:admm}
	\begin{array}{ll}
		\mbox{minimize} & \sum_{s=1}^N f_s(x_s) \\
		\mbox{subject to} & x_s = z, \quad s = 1,\ldots,N, \\
		& z \geq 0.
	\end{array}
\end{equation}
The objective function in Problem (\ref{prob:admm}) is separable across scenarios: the only link between $\{f_s(x_s)\}_{s=1}^N$ is the requirement that $x_1 = \ldots = x_N = z$. However, even with the consensus constraint, ADMM is able to split this problem into independent subproblems. The reader is referred to \cite{BoydParikh:2011} for a detailed description of ADMM.

ADMM starts from an initial estimate of the spot intensity vector $z^{(0)} \in \mathbf{R}_+^n$, the constant parameter $\rho > 0$, and the dual variables $y_s^{(0)} \in \mathbf{R}^n$ for $s = 1,\ldots,N$. (The value $1/\rho$ is known as the step size). In each iteration $k = 0,1,2,\ldots$,
\begin{enumerate}
	\item For $s = 1,\ldots,N$, solve for the scenario spot intensities
	\begin{equation}
		\label{prob:admm_sub}
		x_s^{(k+1)} = \arg\min_{x_s} f_s(x_s) + \frac{\rho}{2}\|x_s - z^{(k)} + y_s^{(k)}/\rho\|_2^2.
	\end{equation}
	\item Project the average of the scenario spot intensities onto the nonnegative orthant:
	\[
	z^{(k+1)} = \max\left(\frac{1}{N} \sum_{s=1}^N x_s^{(k+1)}, 0\right).
	\]
	\item For $s = 1,\ldots,N$, update the parameters
	\begin{equation}
		\label{eq:dual_update}
		y_s^{(k+1)} = y_s^{(k)} + \rho(x_s^{(k+1)} - z^{(k+1)}).
	\end{equation}
	\item Terminate if stopping criterion (\ref{eq:admm_stop}) is satisfied and return $x^{\star} = z^{(k+1)}$ as the solution.
\end{enumerate}
Notice that the scenario subproblems (\ref{prob:admm_sub}) can be solved in parallel.

\paragraph{Solving the subproblems.} Subproblem (\ref{prob:admm_sub}) is an unconstrained least-squares problem with a closed-form solution $x_s^{(k+1)}$, which can be obtained by solving a system of linear equations of the form $B^{(s)}x_s^{(k+1)} = c^{(k)}$, where $B^{(s)} \in \mathbf{R}_+^{n \times n}$ is an iteration-independent symmetric positive definite matrix and $c^{(k)} \in \mathbf{R}_+^n$. This system is sometimes referred to as the normal equations. Since the value of $B^{(s)}$ remains the same throughout the ADMM loop, we can solve Subproblem (\ref{prob:admm_sub}) efficiently by forming the Cholesky factorization of $B^{(s)}$ once prior to the start of ADMM and caching it, then using backward substitution to calculate $x_s^{(k+1)}$ each iteration. This reduces the computational load and ensures our subproblem step is still parallelizable. See Appendix \ref{app:admm_sub} for mathematical details.

\paragraph{ADMM with Barzilai-Borwein step size.} While ADMM will converge for any $\rho > 0$, the value of $\rho$ often has an effect on the practical convergence rate. In some cases, allowing $\rho$ to vary in each iteration will result in faster convergence. One popular method for selecting a variable step size is the Barzilai-Borwein (BB) method \cite{BarzilaiBorwein:1988,Raydan:1997,DaiFletcher:2005}. This method has shown great success when combined with ADMM \cite{ZarepishehXing:2018}. The BB step size in iteration $k$ is given by
\begin{equation}
	\label{eq:bb_step}
	\beta^{(k)} = -\frac{\Delta d^{(k)T} \Delta y^{(k)}}{\|\Delta d^{(k)}\|_2^2},
\end{equation}
where $\Delta y^{(k)} = y^{(k)} - y^{(k-1)}$ and $\Delta d^{(k)} = (x^{(k+1)} - z^{(k+1)}) - (x^{(k)} - z^{(k)})$. If $\beta^{(k)} > 0$, we use it in place of $\rho$ in our dual variable update \ref{eq:dual_update}. Otherwise, we revert to the user-defined constant $\rho > 0$.

Note that Subproblem (\ref{prob:admm_sub}) always uses the default $\rho$, since for speed purposes, we do not recompute the Cholesky factorization each iteration. 
In addition, the BB step size is calculated using simple linear algebra, not the serial loop required by a backtracking line search (e.g., \textit{Armijo}). As we will see in Section \ref{sec:results}, this gives ADMM a significant advantage over PGD in terms of runtime and memory efficiency. 

\paragraph{Stopping criterion.} \replaced[id=R1]{The convergence of ADMM with variable step size is still an active topic of research \cite{XuGoldstein:2017,XuYang:2017,BotCsetnek:2019}. In our algorithm, we employ a stopping criterion based on the optimality conditions of Problem (\ref{prob:admm}), which we have observed works well in practice. This translates into checking whether the residuals associated with these conditions,}{Since Problem (\ref{prob:nnls_full}) is convex, ADMM is guaranteed to converge to a solution \mbox{\cite{BoydParikh:2011,Eckstein:1992,Gabay:1983}}. The difference between the spot intensities found in each scenario will converge to zero, so the constraint $x_s = z$ for $s = 1,\ldots,N$ is met at the optimum. As a consequence, the residuals associated with this constraint,}
\begin{align*}
	r_{\text{prim}}^{(k)} &= (x_1^{(k)} - z^{(k)}, \ldots, x_N^{(k)} - z^{(k)}), \\
	r_{\text{dual}}^{(k)} &= (\rho(z^{(k)} - z^{(k-1)}), \ldots, \rho(z^{(k)} - z^{(k-1)})),
\end{align*}
\replaced[id=R1]{are close to zero.}{will also converge to zero.} (Here $r_{\text{prim}}$ is referred to as the primal residual and $r_{\text{dual}}$ as the dual residual). Thus, a reasonable stopping criterion is
\begin{equation}
	\label{eq:admm_stop}
	\|r_{\text{prim}}^{(k)}\|_2 \leq \epsilon_{\text{prim}}, \quad \|r_{\text{dual}}^{(k)}\|_2 \leq \epsilon_{\text{dual}}
\end{equation}
for tolerances $\epsilon_{\text{prim}} > 0$ and $\epsilon_{\text{dual}} > 0$. These tolerances are typically chosen with respect to user-defined cutoffs $\epsilon_{\text{abs}} > 0$ and $\epsilon_{\text{rel}} > 0$ using
\begin{align*}
	\epsilon_{\text{prim}} &= \epsilon_{\text{abs}}\sqrt{Nn} +  \epsilon_{\text{rel}}\max(\|x^{(k)}\|_2, \sqrt{N}\|z^{(k)}\|_2), \\
	\epsilon_{\text{dual}} &= \epsilon_{\text{abs}}\sqrt{Nn} + \epsilon_{\text{rel}}\|y^{(k)}\|_2,
\end{align*}
where $x^{(k)} := (x_1^{(k)}, \ldots, x_N^{(k)})$ and $y^{(k)} := (y_1^{(k)}, \ldots, y_N^{(k)})$. \added[id=R2]{Intuitively, we want to stop when 1) the scenario-specific spots are approximately equal ($r_{\text{prim}}^{(k)} \approx 0$ implies $x_1^{(k)} \approx \ldots \approx x_N^{(k)} \approx z^{(k)}$) and 2) the most recent iteration of ADMM yielded little change in the spot intensities ($r_{\text{prim}}^{(k)} \approx 0$ implies $z^{(k)} \approx z^{(k-1)}$). This indicates that the optimality conditions for Problem (\ref{prob:admm}) have been fulfilled.} We refer the reader to \cite[Section 3.3]{BoydParikh:2011} for a derivation of the stopping criterion.

\subsection{Patient population and computational framework}
\label{sec:patients}

We tested our ADMM algorithm with BB step size (ADMM-BB) on four head-and-neck cancer patient cases from The Cancer Imaging Archive (TCIA) \cite{TCIA:paper,TCIA:data}. For each patient, only the primary clinical target volume (CTV) was considered in the optimization, with a prescribed dose of $D_{\text{pres}} = 70$ Gy delivered in $35$ fractions of $2$ Gy per fraction. Dose calculations were performed in MATLAB using the open-source software MatRad and the pencil beam dose calculation algorithm \cite{MatRad,MatRad:Int}. The proton spots were placed on a rectangular grid, covering the planning target volume (PTV) region plus $1$ mm out from its perimeter with a spot spacing of $5$ mm. All patients were planned using two co-planar beams (see Table \ref{table:patient_info} for details).

To create uncertainty scenarios, we simulated range over- and under-shoots by rescaling the stopping power ratio (SPR) image $\pm 3.5\%$ following typical range margin recipes used in proton therapy \cite{TaastiRichter:2018,Paganetti:2012}. We also simulated setup errors by shifting the isocenter $\pm 3$ mm in the $x$, $y$, and $z$ direction. Combining the range and setup errors gave us a total of $13$ scenarios, including the nominal scenario. For each scenario, we computed the dose-influence matrix using a relative biological effectiveness (RBE) of $1.1$. 

We implemented PGD and ADMM-BB in Python using the built-in \texttt{multiprocessing} library. This library supports parallel computation across multiple CPUs and CPU threads/cores. The algorithms were executed on a server with $2$ Intel Xeon Gold 6248 CPUs @ $2.50$ GHz / $20$ cores and $128$ GB RAM. We terminated PGD when the relative change in the objective value was about $10^{-4}$ to $10^{-5}$. For ADMM-BB, we combined this same criterion with the residual stopping criterion \ref{eq:admm_stop} using $\epsilon_{\text{abs}} = \epsilon_{\text{rel}} = 10^{-6}$.

\begin{table}
	\centering
	\begin{tabular}{c|cccc}
		\hline \hline
		& \multicolumn{4}{|c}{Patient} \\
		\cline{2-5}
		&  1 & 2 & 3 & 4 \\
		\hline
		Beam configuration & $40^{\circ}, 320^{\circ}$ & $220^{\circ}, 320^{\circ}$ & $30^{\circ}, 60^{\circ}$ & $30^{\circ}, 300^{\circ}$ \\ 
		CTV volume (cm$^3$) & 91.4 & 2.8 & 72.9 & 59.5 \\
		PTV volume (cm$^3$) & 162.7 & 12.4 & 129.9 & 107.5 \\
		Number of voxels ($m$) & 98901 & 50728 & 119258 & 126072 \\
		Number of spots ($n$) & 5762 & 707 & 4697 & 3737 \\
		\hline \hline
	\end{tabular}
	\caption{Head-and-neck cancer patient information.}
	\label{table:patient_info}
\end{table}

\section{Results}
\label{sec:results}

\subsection{Algorithm runtime comparisons}
\label{sec:runtimes}

\begin{figure}
	\centering
	\begin{subfigure}[b]{0.48\textwidth}
		\caption{Patient 1}
		\includegraphics[width=\textwidth]{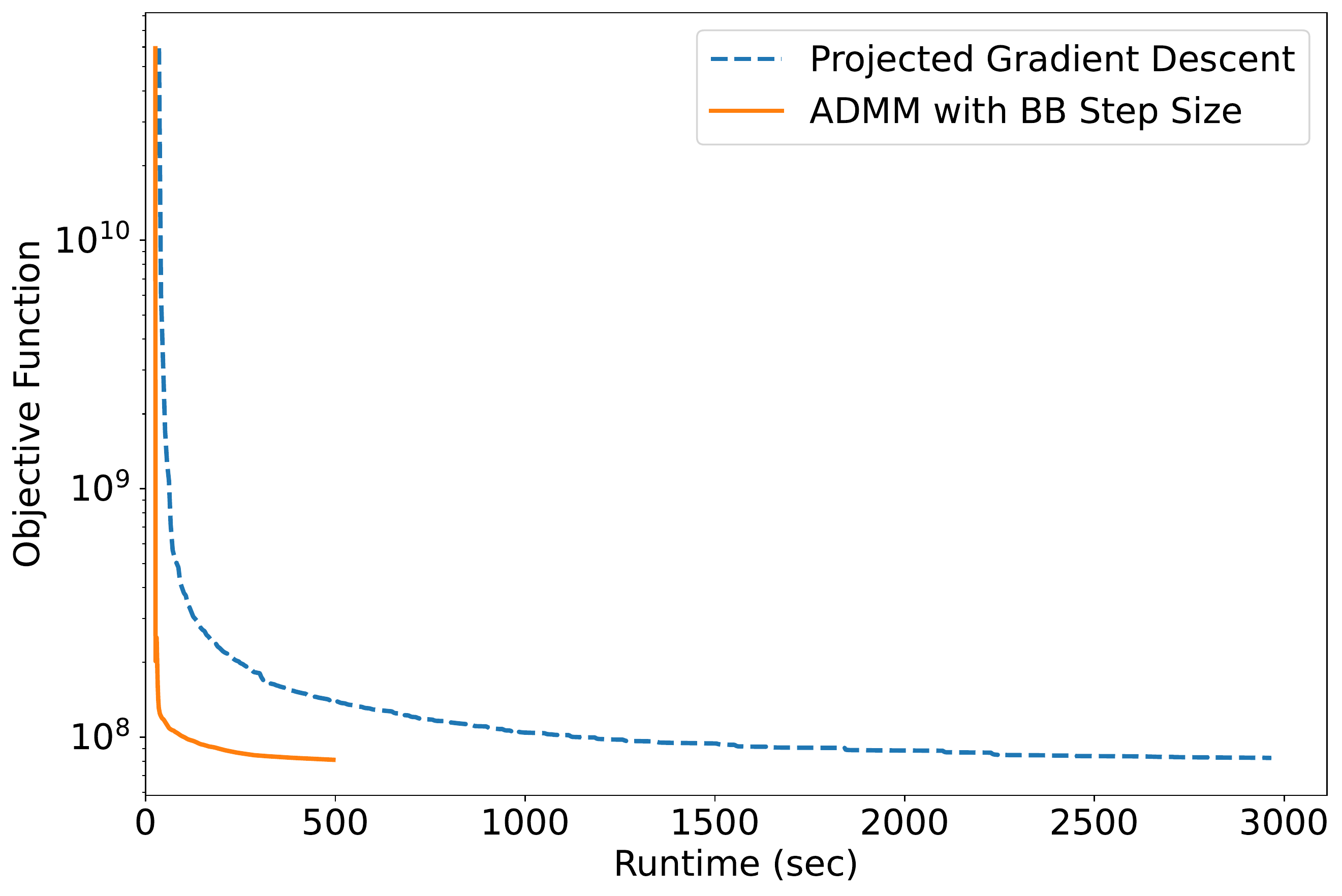}
	\end{subfigure}
	\begin{subfigure}[b]{0.48\textwidth}
		\caption{Patient 2}
		\includegraphics[width=\textwidth]{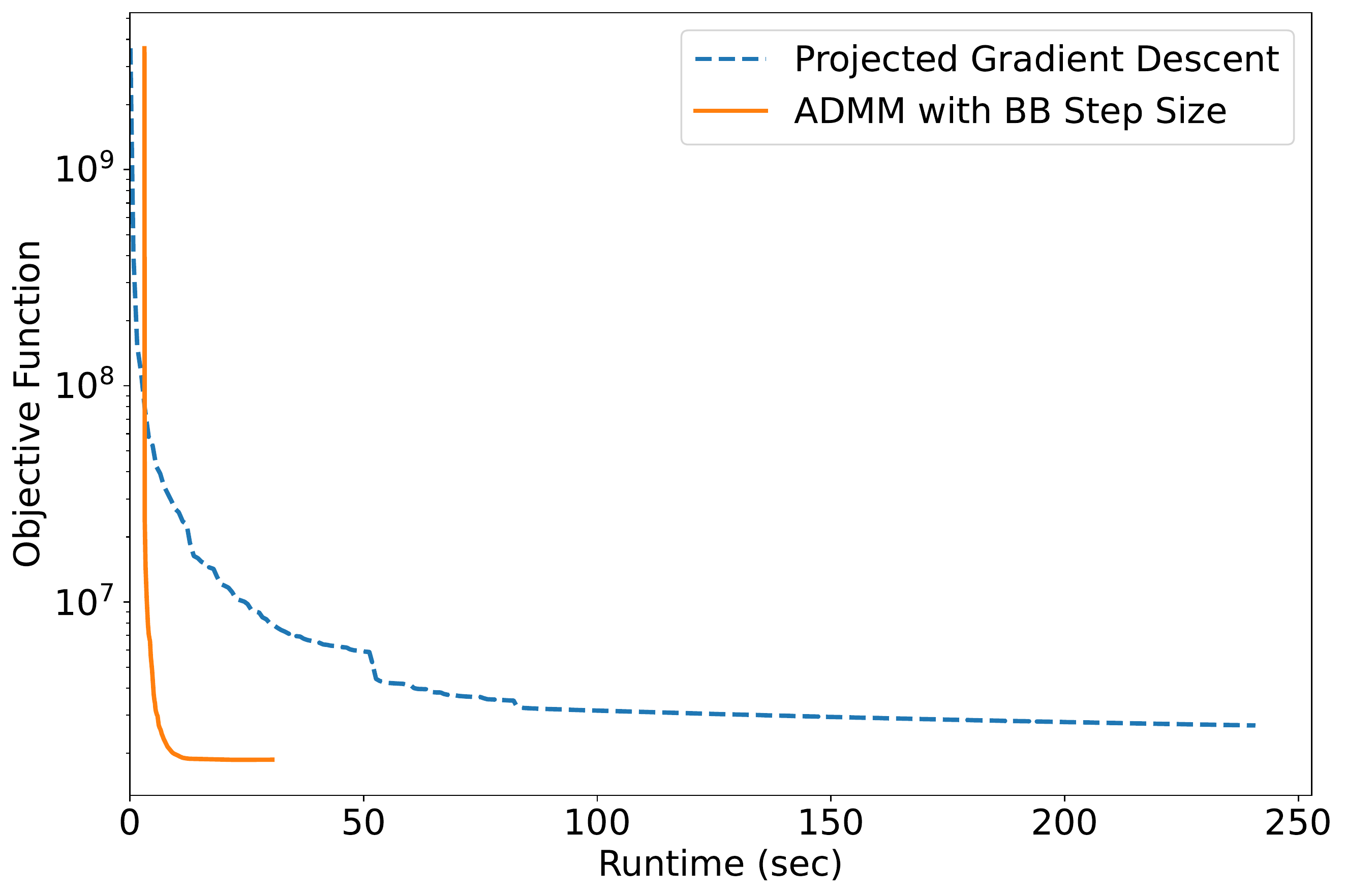}
	\end{subfigure} 
	\\ \medskip
	\begin{subfigure}[b]{0.48\textwidth}
		\caption{Patient 3}
		\includegraphics[width=\textwidth]{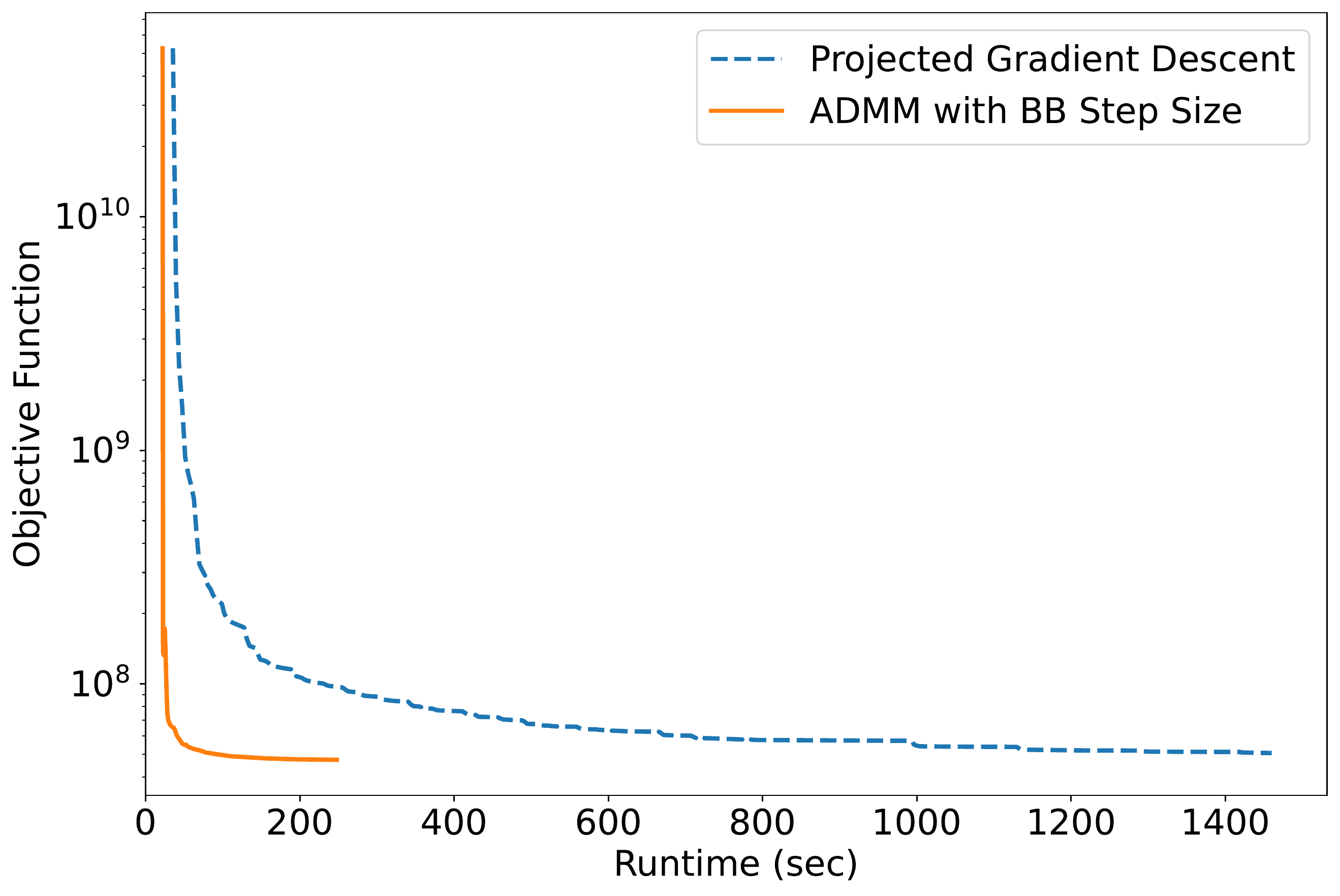}
	\end{subfigure}
	\begin{subfigure}[b]{0.48\textwidth}
		\caption{Patient 4}
		\includegraphics[width=\textwidth]{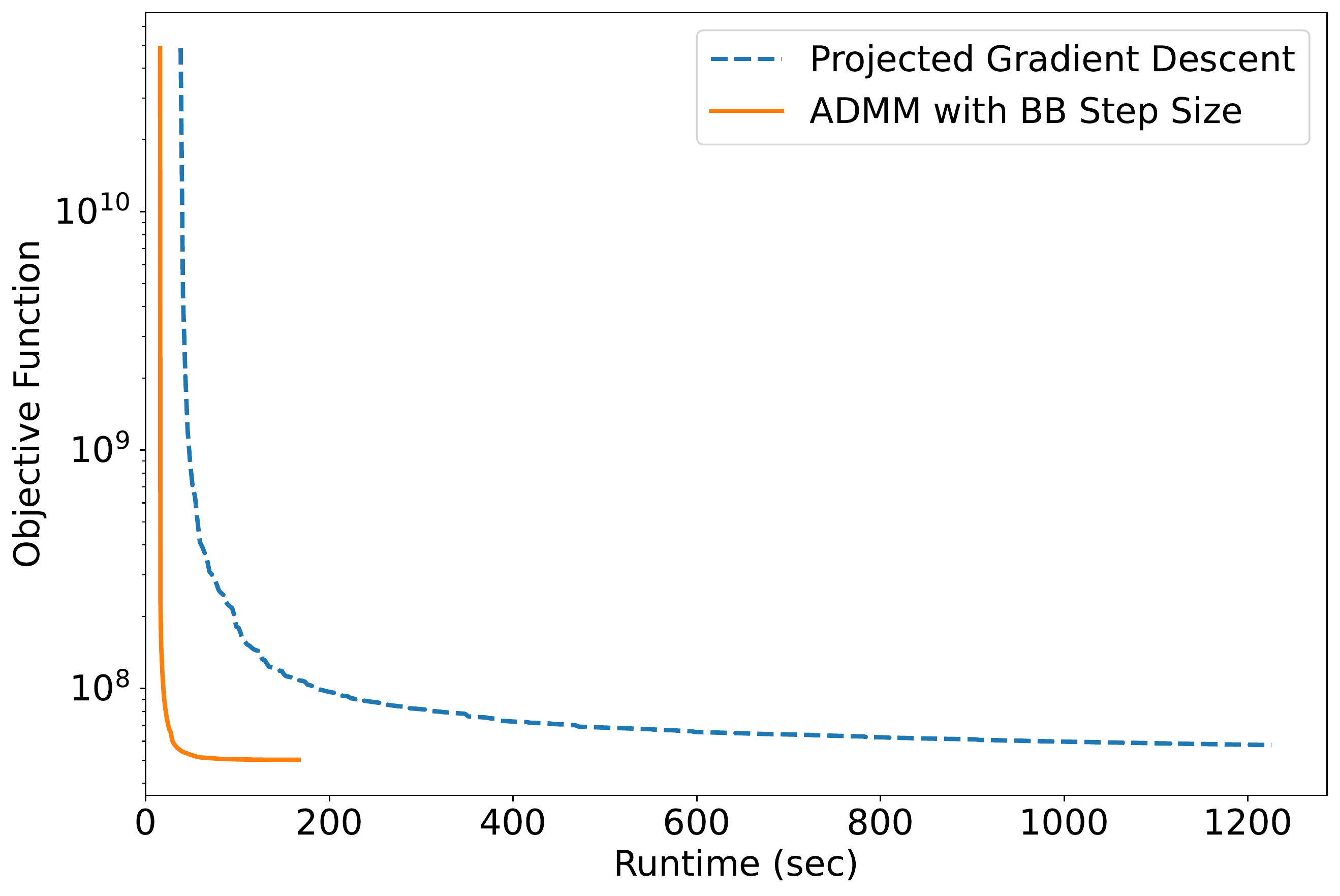}
	\end{subfigure}
	\caption{Objective function value versus algorithm runtime (sec) for all patient cases.}
	\label{fig:objective_times}
\end{figure}

Figure \ref{fig:objective_times} shows the objective function value over the course of the total runtime of PGD and ADMM-BB, for all four patients. Clearly for all patients, ADMM-BB converges much faster than PGD to approximately the same objective value. On patient $1$, a relatively large case, ADMM-BB took $8.3$ minutes to converge, whereas PGD required $50$ minutes to achieve roughly the same objective. For a smaller case like patient $2$, ADMM-BB converged in $31$ seconds, while PGD required $241$ seconds. Altogether across all patients, ADMM-BB converged on average {\bf 6 to 7 times faster} than PGD to a very similar objective and spot intensities --- a significant speedup.

\subsection{Treatment plan comparisons}
\label{sec:plan_comp}

\begin{figure}
	\centering
	\includegraphics[width=0.95\textwidth]{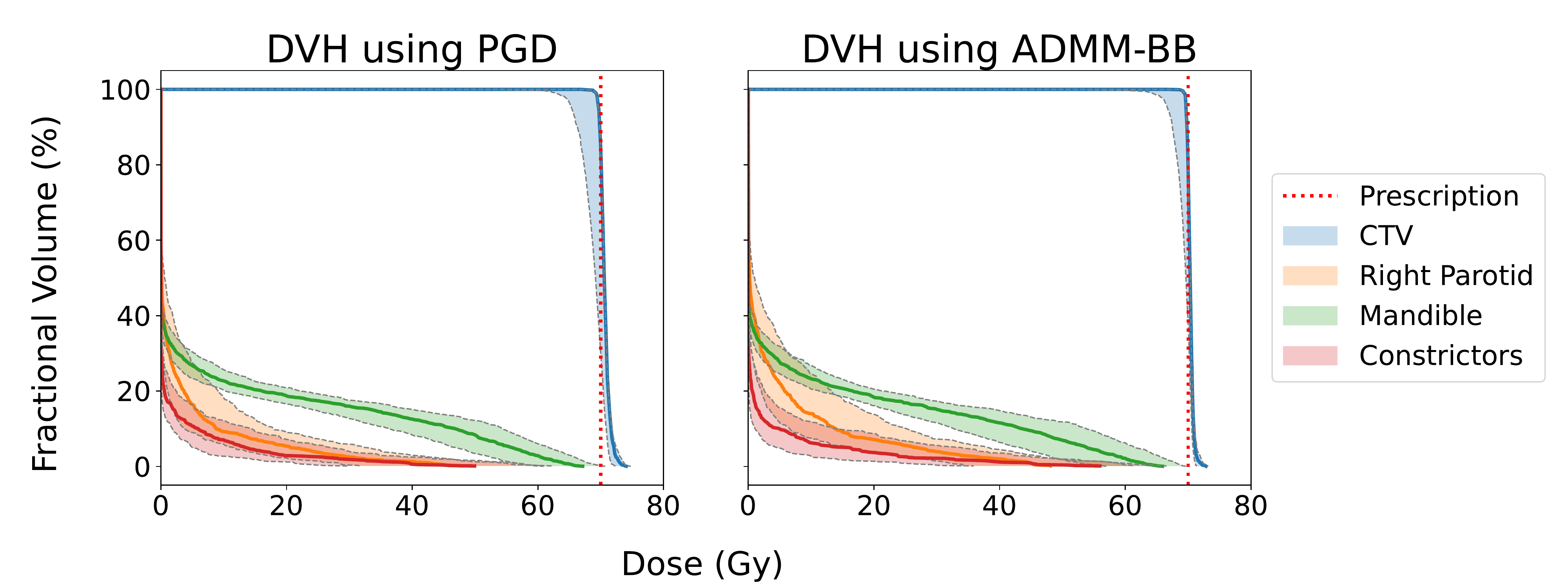}
	\caption{Dose-volume histogram (DVH) bands across all scenarios for patient $2$.}
	\label{fig:dvh_bands-p4}
\end{figure}

\begin{figure}
	\centering
	\includegraphics[width=0.95\textwidth]{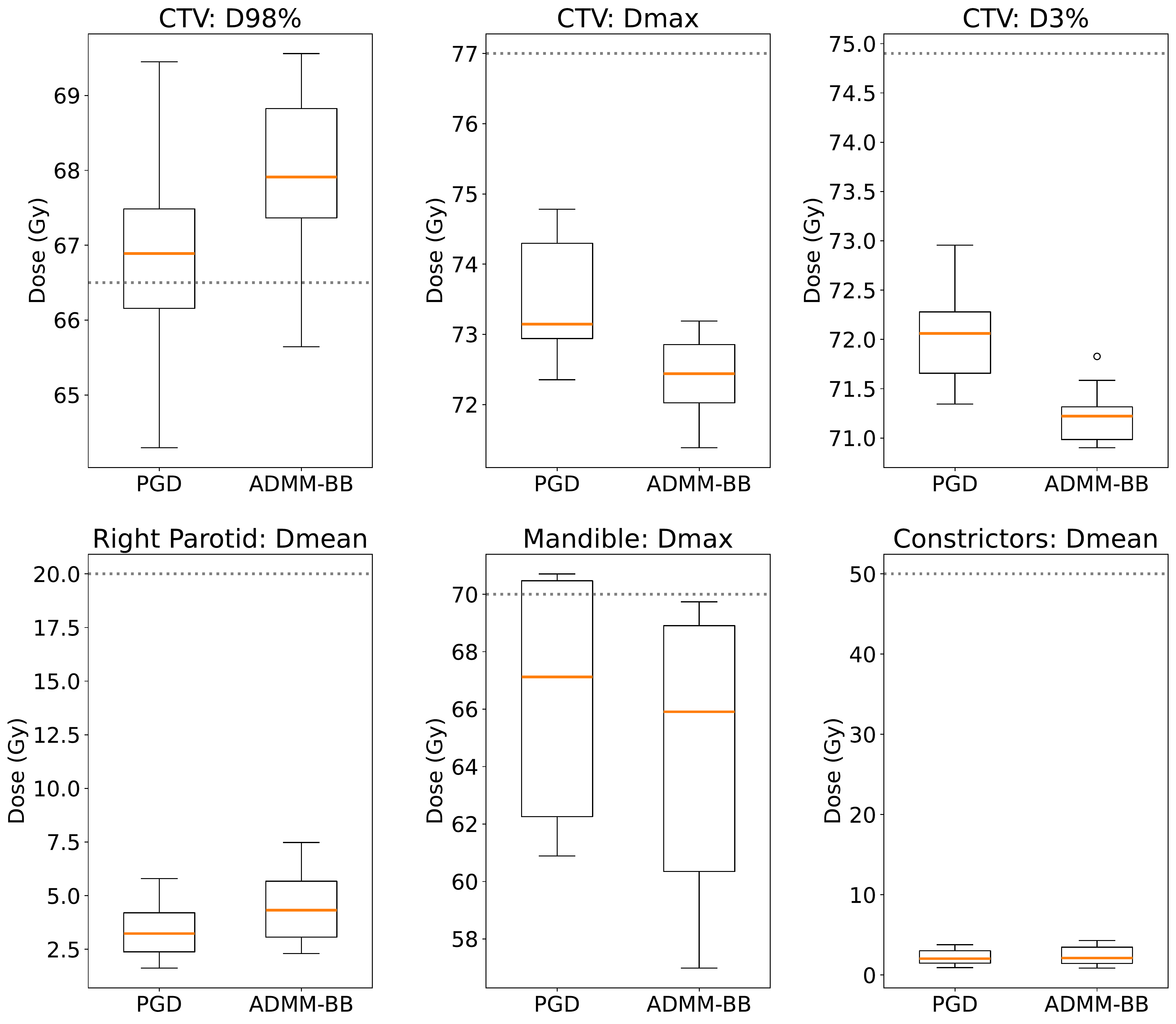}
	\caption{Dose-volume clinical metrics for patient $2$. The box plot spans the values over the $13$ uncertainty scenarios, with the orange line indicating the median. The dotted lines mark the clinical constraints: higher is better for D$98\%$ on the CTV, while lower is better for all other metrics.}
	\label{fig:clinical_metrics-p4}
\end{figure}

Figures \ref{fig:dvh_bands-p4} and  \ref{fig:clinical_metrics-p4} compare the treatment plans produced by PGD and ADMM-BB for patient $2$. Figure \ref{fig:dvh_bands-p4} depicts the DVH bands: each band, color-coded to a particular structure, represents the range of DVH curves across all $13$ scenarios, while the corresponding solid curve is the DVH in the nominal scenario. The vertical dotted line marks the prescription $D_{\text{pres}} = 70$ Gy. It is clear from an inspection of the figure that PGD and ADMM-BB converge to near-identical DVH bands. The only difference is that ADMM-BB achieves a tighter CTV band around $D_{\text{pres}}$ at the expense of a slight increase in the mean dose to the right parotid (Figure \ref{fig:dvh_bands-p4}). Figure \ref{fig:clinical_metrics-p4} shows the box plots of a few dose-volume clinical metrics for this patient. The box plots depict the value of the metric for each of the $13$ uncertainty scenarios. The dotted line represents the clinical constraint for the metric; this line is a lower bound for D$98\%$ on the CTV and an upper bound for all other metrics. (See Table \ref{table:dose_constrs} for a full description of the criteria). It is apparent from the plots that both PGD and ADMM-BB produce robust plans that respect the clinical constraints in nearly all scenarios. On the OAR metrics, PGD and ADMM-BB achieve similar performance: the mean doses to the right parotid and the constrictors are nearly identical, while the maximum dose to the mandible is slightly lower for ADMM-BB. However, ADMM-BB outperforms PGD on the CTV metrics.

\begin{table}
	\centering
	\begin{tabular}{c|c}
		\hline \hline
		Structure & Dose constraint \\
		\hline
		CTV & \makecell{$\text{D}98\% > 66.5$ Gy \\ $\text{Dmax} < 77$ Gy \\ $\text{D}3\% < 74.9$ Gy} \\
		\hline
		Parotid & $\text{Dmean} < 20$ Gy \\
		Mandible & $\text{Dmax} < 70$ Gy \\
		Larynx & $\text{Dmean} < 45$ Gy \\
		Constrictors & $\text{Dmean} < 50$ Gy \\
		\hline \hline
	\end{tabular}
	\caption{Recommended clinical dose bounds for head-and-neck cancer patients \cite{Taasti:Hierarchical}.}
	\label{table:dose_constrs}
\end{table}

\begin{figure}
	\centering
	\includegraphics[width=0.95\textwidth]{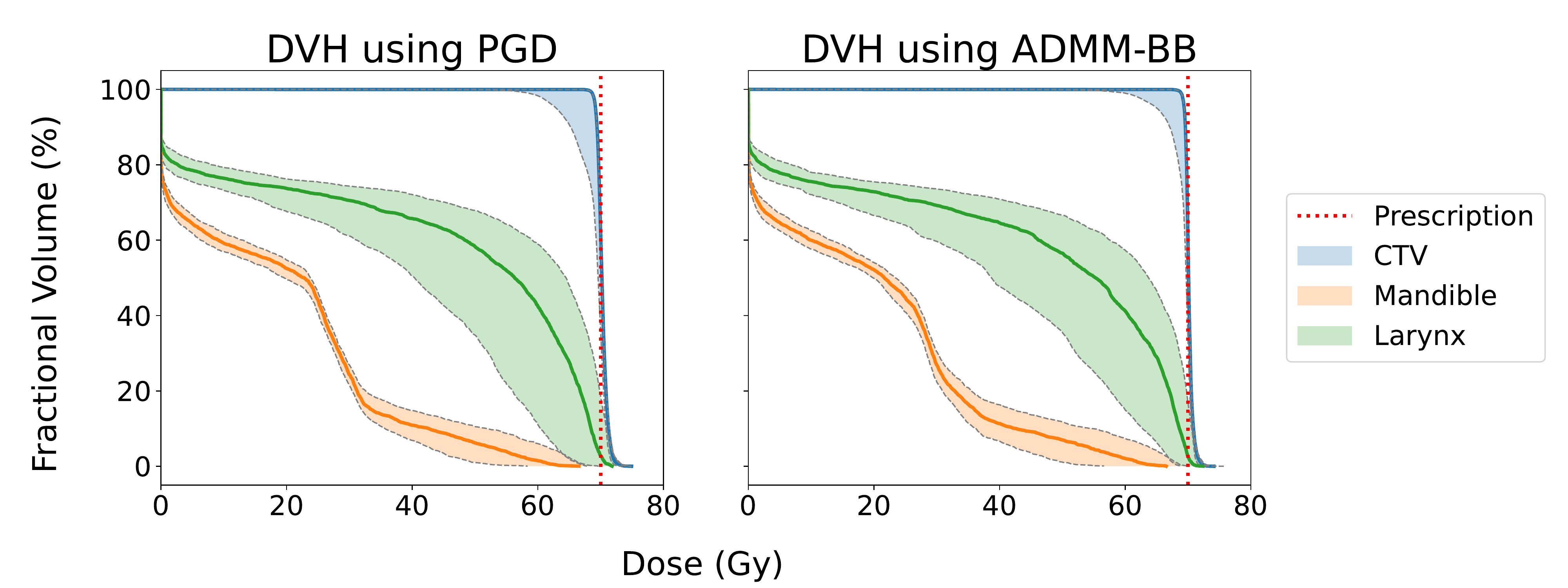}
	\caption{Dose-volume histogram (DVH) bands across all scenarios for patient $4$.}
	\label{fig:dvh_bands-p5}
\end{figure}

\begin{figure}
	\centering
	\includegraphics[width=0.95\textwidth]{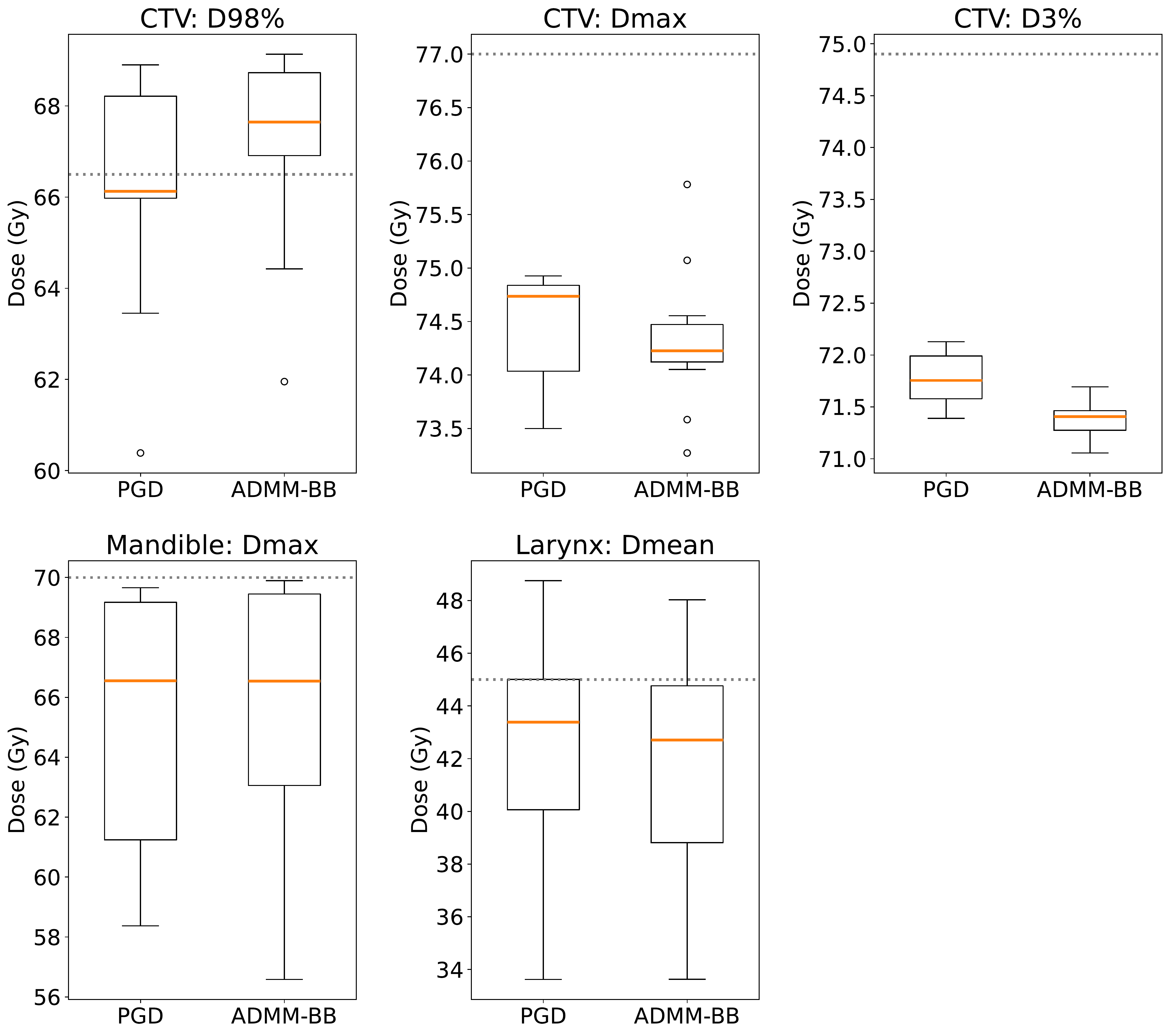}
	\caption{Dose-volume clinical metrics for patient $4$. The box plot spans the values over the $13$ uncertainty scenarios, with the orange line indicating the median. The dotted lines mark the recommended clinical constraints: higher is better for D$98\%$ on the CTV, while lower is better for all other metrics.}
	\label{fig:clinical_metrics-p5}
\end{figure}

Figures \ref{fig:dvh_bands-p5} and \ref{fig:clinical_metrics-p5} depict the DVH bands and clinical metrics for patient $4$. Again, the PGD and ADMM-BB treatment plans are nearly identical. ADMM-BB produces a tighter CTV band below $70$ Gy, indicating superior robustness, at the expense of a somewhat wider DVH band on the mandible. The box plots of the dose-volume metrics for the ADMM-BB plan are also largely within the desired clinical constraints, and the CTV box plots in particular show an improvement over those of PGD (Figure \ref{fig:dvh_bands-p5}), similar to what we saw in patient $2$. The PGD and ADMM-BB treatment plans for other patients reflect the same pattern; we have omitted them here for brevity. Overall, we see that ADMM-BB produces plans identical to or better than the plans produced by PGD, but in a fraction of the runtime.

\subsection{Scalability over multiple scenarios}
\label{sec:scaling}

\begin{figure}
	\centering
	\includegraphics[width=0.95\textwidth]{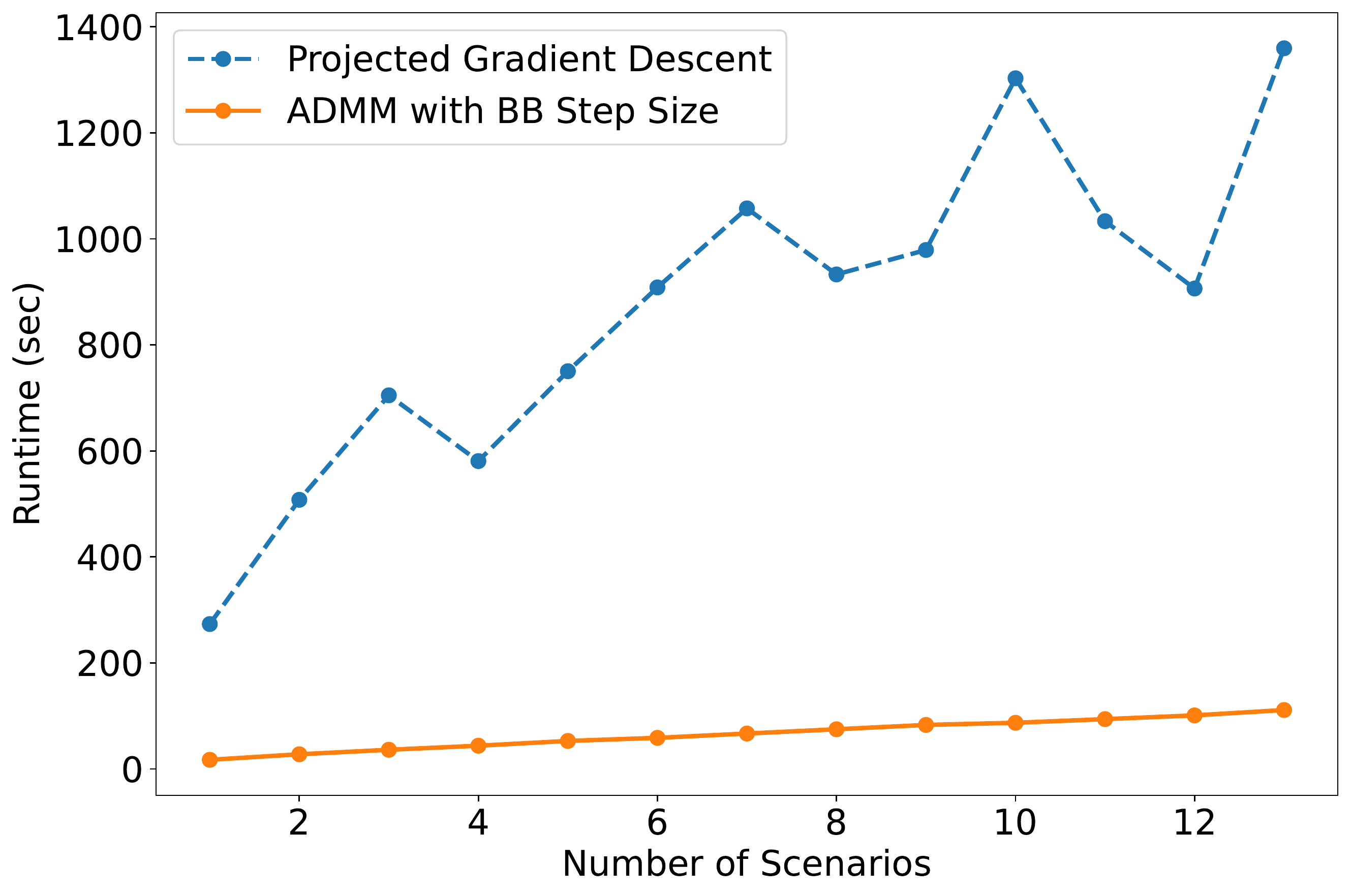}
	\caption{Algorithm runtime (sec) versus number of scenarios ($N$) for patient 4's treatment plan.}
	\label{fig:times-scenario_scaling}
\end{figure}

Not only is ADMM-BB faster than PGD, it also scales better with the number of scenarios. Figure \ref{fig:times-scenario_scaling} shows the time required by PGD and ADMM-BB to produce a treatment plan for patient $4$ for different numbers of scenarios in the optimization problem. Specifically, we solved Problem (\ref{prob:nnls_full}) for $N \in \{1,2,\ldots,13\}$ and recorded the total runtime in each case. Parallelization in ADMM-BB was carried out across the included scenarios. It is clear from the figure that ADMM-BB outperforms PGD by a wide margin. ADMM-BB's runtime increases modestly from $17.3$ seconds at $N = 1$ to $111.2$ seconds at $N = 13$. By contrast, PGD goes from $273.4$ seconds at $N = 1$ to $1359.4$ seconds at $N = 13$. At its peak, ADMM-BB is over {\bf 12 times faster} than PGD. This pattern holds true for all the other patients as well.

\section{Discussion}
\label{sec:discussion}

In this study, we presented a distributed optimization method for solving robust proton treatment planning problems. Our method splits the problem into smaller parts, which can be handled on separate machines/processors, allowing us to reduce the overall planning time and overcome the limits (e.g., memory) of single-machine computation. As a result, we can efficiently plan large patient cases with many voxels and spots. We can also consider more uncertainty scenarios for any given patient, enabling us to produce more robust treatment plans.

We have used PGD as our benchmark for comparison because it is commonly employed to solve unconstrained treatment planning problems \added[id=R1]{\cite{Bortfeld:1999,HristovStavrev:2002,GhobadiGhaffari:2012,YaoTempleton:2014}}. More advanced versions of gradient descent, such as the \added[id=R1]{nonlinear} conjugate gradient (CG) method and Nesterov-accelerated gradient descent, can also be applied to Problem (\ref{prob:nnls_full}). While these algorithms may work better than PGD, they possess the same dependence on the line search step \added[id=R1]{\cite{NocedalWright:2006,GonzagaKaras:2013}}, which was the main source of computational burden in our experiments. A variant of PGD exists that uses the BB step size \cite{KimSra:2013}. The removal of line search does speed up PGD, but experiments show that ADMM-BB still generally converges faster in a non-distributed setting \cite{ZarepishehXing:2018}. In sum, both the ADMM subproblem decomposition and the BB step size combine to give ADMM-BB a clear advantage over gradient descent.

Another advantage of ADMM is that it supports a wide variety of constraints. While PGD can only handle box constraints (i.e., upper and lower bounds on the spot intensities), ADMM can easily be modified to handle any type of linear constraint, such as maximum and mean dose constraints. More advanced versions of ADMM can also handle nonlinear constraints \cite{GaoCurtis:2020,LatorreCevher:2019,BenningKnoll:2015}. Other optimization algorithms exist that support linear/nonlinear constraints (e.g., interior point method \cite{BoydVandenberghe:2004}, sequential quadratic programming \cite{NocedalWright:2006}), but generally, they require the calculation of the Hessian matrix, which is computationally expensive for robust treatment planning problems.  

In our formulation, we have chosen to split Problem (\ref{prob:admm_sub}) by scenarios. It is also possible to split the problem by voxels, e.g., create a separate subproblem for each target and organ-at-risk. Indeed, there are multi-block versions of ADMM that allow arbitrary splitting along rows and columns \cite{ParikhBoyd:2014,SunYe:2015,DengYin:2017,MihicYe:2021}, so we can theoretically accommodate any partition of voxels/spots. This is a natural direction for future research. 

Finally, we plan to extend the implementation of ADMM-BB to other platforms. Our current software implementation parallelizes computation across multiple CPUs and CPU cores. We intend to add support for parallelization across GPUs, which should produce further speed improvements. With the rise of cloud computing, data storage and processing is increasingly moving to remote high performance computing (HPC) clusters. Our long-term goal is to implement ADMM-BB on these clusters, so that each subproblem (and its corresponding portion of the data) is handled by a separate machine or group of machines in the cluster. This will allow us to solve treatment planning problems of immense size. We foresee ADMM-BB's application in many data-intensive treatment environments, such as beam angle optimization and volumetric modulated arc therapy (VMAT), as well as more recent treatment planning modalities like 4$\pi$ \cite{DongLee:2013} and station parameter optimized radiation therapy (SPORT) \cite{ZarepishehXing:2015}.

\section{Conclusion}
\label{sec:conclusion}

We have developed a fast, distributed method for robust proton treatment planning. Our method splits the treatment planning problem into smaller subproblems, which can then be solved in parallel, improving runtime and memory efficiency. Moreover, we showed that the method scales well with the number of uncertainty scenarios. These advantages allow us to 1) shorten the time required by the treatment planning process, 2) incorporate more uncertainty scenarios into the robust optimization problem, and 3) improve plan quality by exploring a larger space of parameters.

\bibliographystyle{alpha}
\bibliography{robust_proton}

\newcommand{\etalchar}[1]{$^{#1}$}
\begin{thebibliography}{WWG{\etalchar{+}}18}

\bibitem[BB88]{BarzilaiBorwein:1988}
J.~Barzilai and J.~M. Borwein.
\newblock Two-point step size gradient methods.
\newblock {\em {IMA} Journal of Numerical Analysis}, 8(1):141--148, 1988.

\bibitem[BC19]{BotCsetnek:2019}
R.~I. Bo\c{t} and E.~R. Csetnek.
\newblock {ADMM} for monotone operations: Convergence analysis and rates.
\newblock {\em Advances in Computational Mathematics}, 45(1):327--359, 2019.

\bibitem[BKSV15]{BenningKnoll:2015}
M.~Benning, F.~Knoll, C.-B. Schonlieb, and T.~Valkonen.
\newblock Preconditioned {ADMM} with nonlinear operator constraint.
\newblock In {\em {IFIP} Conference on System Modeling and Optimization}, pages
  117--126, 2015.

\bibitem[Bor99]{Bortfeld:1999}
T.~Bortfeld.
\newblock Optimized planning using physical objectives and constraints.
\newblock {\em Seminars in Radiation Oncology}, 9(1):20--34, January 1999.

\bibitem[BPC{\etalchar{+}}11]{BoydParikh:2011}
S.~Boyd, N.~Parikh, E.~Chu, B.~Peleato, and J.~Eckstein.
\newblock Distributed optimization and statistical learning via the alternating
  direction method of multipliers.
\newblock {\em Foundations and Trends in Machine Learning}, 3(1):1--122, 2011.

\bibitem[BT89]{BertsekasTsitsiklis:1989}
D.~P. Bertsekas and J.~N. Tsitsiklis.
\newblock {\em Parallel and Distributed Computation: Numerical Methods}.
\newblock Prentice-Hall, 1989.

\bibitem[BV04]{BoydVandenberghe:2004}
S.~Boyd and L.~Vandenberghe.
\newblock {\em Convex Optimization}.
\newblock Cambridge University Press, 2004.

\bibitem[CVS{\etalchar{+}}13]{TCIA:paper}
K.~Clark, B.~Vendt, K.~Smith, J.~Freymann, J.~Kirby, P.~Koppel, S.~Moore,
  S.~Phillips, D.~Maffitt, M.~Pringle, L.~Tarbox, and F.~Prior.
\newblock The cancer imaging archive ({TCIA}): Maintaining and operating a
  public information repository.
\newblock {\em Journal of Digital Imaging}, 26(6):1045--1057, 2013.

\bibitem[DCRS15]{DharYi:2015}
S.~Dhar, Y.~Congrui, N.~Ramakrishnan, and M.~Shah.
\newblock {ADMM} based machine learning on {Spark}.
\newblock In {\em {IEEE} Conference on Big Data}, pages 1174--1182, 2015.

\bibitem[DF05]{DaiFletcher:2005}
Y.H. Dai and R.~Fletcher.
\newblock Projected {B}arzilai-{B}orwein methods for large-scale
  box-constrained quadratic programming.
\newblock {\em Numerische Mathematik}, 100(1):21--47, 2005.

\bibitem[DHS11]{DuchiHazan:2011}
J.~Duchi, E.~Hazan, and Y.~Singer.
\newblock Adaptive subgradient methods for online learning and stochastic
  optimization.
\newblock {\em Journal of Machine Learning Research}, 12(61):2121--2159, 2011.

\bibitem[DLPY17]{DengYin:2017}
W.~Deng, M.-J. Lai, Z.~Peng, and W.~Yin.
\newblock Parallel multi-block {ADMM} with $o(1/k)$ convergence.
\newblock {\em Journal of Scientific Computing}, 71(1):712--736, 2017.

\bibitem[DLR{\etalchar{+}}13]{DongLee:2013}
P.~Dong, P.~Lee, D.~Ruan, T.~Long, E.~Romeijn, D.~A. Low, P.~Kupelian,
  J.~Abraham, Y.~Yang, and K.~Sheng.
\newblock $4\pi$ noncoplanar stereotactic body radiation therapy for centrally
  located or larger lung tumors.
\newblock {\em International Journal of Radiation Oncology Biology Physics},
  86(3):407--413, 2013.

\bibitem[FHLY15]{FangHe:2015}
E.~X. Fang, B.~He, H.~Liu, and X.~Yuan.
\newblock Generalized alternating direction method of multipliers: New
  theoretical insights and applications.
\newblock {\em Mathematical Programming Computation}, 7(2):149--187, June 2015.

\bibitem[FXB22]{FuBoyd:2022}
A.~Fu, L.~Xing, and S.~Boyd.
\newblock Operator splitting for adaptive radiation therapy with nonlinear
  health constraints.
\newblock {\em Optimization Methods and Software}, pages 1--24, June 2022.
\newblock
  \href{https://doi.org/10.1080/10556788.2022.2078824}{doi:10.1080/10556788.2022.2078824}.

\bibitem[GFK{\etalchar{+}}18]{GuFan:2018}
Y.~Gu, J.~Fan, L.~Kong, S.~Ma, and H.~Zou.
\newblock {ADMM} for high-dimensional sparse penalized quantile regression.
\newblock {\em Technometrics}, 60(3):319--331, 2018.

\bibitem[GGA{\etalchar{+}}12]{GhobadiGhaffari:2012}
K.~Ghobadi, H.~R. Ghaffari, D.~M. Aleman, D.~A. Jaffray, and M.~Ruschin.
\newblock Automated treatment planning for a dedicated multi-source
  intracranial radiosurgery treatment unit using projected gradient and
  grassfire algorithms.
\newblock {\em Medical Physics}, 39(6):3134--3141, June 2012.

\bibitem[GGC20]{GaoCurtis:2020}
W.~Gao, D.~Goldfarb, and F.~E. Curtis.
\newblock {ADMM} for multiaffine constrained optimization.
\newblock {\em Optimization Methods and Software}, 35(2):257--303, 2020.

\bibitem[GK13]{GonzagaKaras:2013}
C.~C. Gonzaga and E.~W. Karas.
\newblock Fine tuning {N}esterov's steepest descent algorithm for
  differentiable convex programming.
\newblock {\em Mathematical Programming}, 138(1):141--166, 2013.

\bibitem[GM76]{GabayMercier:1976}
D.~Gabay and B.~Mercier.
\newblock A dual algorithm for the solution of nonlinear variational problems
  via finite element approximations.
\newblock {\em Computers and Mathematics with Applications}, 2(1):17--40, 1976.

\bibitem[HHG{\etalchar{+}}19]{HuangHu:2019}
Z.~Huang, R.~Hu, Y.~Guo, E.~Chan-Tin, and Y.~Gong.
\newblock {DP-ADMM}: {ADMM}-based distributed learning with differential
  privacy.
\newblock {\em {IEEE} Transactions on Information Forensics and Security},
  15(1):1002--1012, 2019.

\bibitem[HSSF02]{HristovStavrev:2002}
D.~Hristov, P.~Stavrev, E.~Sham, and B.~G. Fallone.
\newblock On the implementation of dose-volume objectives in gradient
  algorithms for inverse treatment planning.
\newblock {\em Medical Physics}, 29(5):848--856, 2002.

\bibitem[KB14]{KingmaBa:2014}
D.~P. Kingma and J.~Ba.
\newblock Adam: A method for stochastic optimization, 2014.
\newblock arXiv:1412.6980.

\bibitem[KSD13]{KimSra:2013}
D.~Kim, S.~Sra, and I.~S. Dhillon.
\newblock A non-monotonic method for large-scale non-negative least squares.
\newblock {\em Optimization Methods and Software}, 28(5):1012--1039, 2013.

\bibitem[LEC19]{LatorreCevher:2019}
F.~Latorre, A.~Eftekhari, and V.~Cevher.
\newblock Fast and provable {ADMM} for learning with generative priors.
\newblock In {\em Advances in Neural Information Processing Systems ({NIPS})
  32}, 2019.

\bibitem[MZY21]{MihicYe:2021}
K.~Mihi{\'c}, M.~Zhu, and Y.~Ye.
\newblock Managing randomization in the multi-block alternating direction
  method of multipliers for quadratic optimization.
\newblock {\em Mathematical Programming Computation}, 13(1):339--413, 2021.

\bibitem[NL18]{NedicLiu:2018}
A.~Nedi\'{c} and J.~Liu.
\newblock Distributed optimization for control.
\newblock {\em Annual Review of Control, Robotics, and Autonomous Systems},
  1:77--103, 2018.

\bibitem[Nol]{TCIA:data}
T.~Nolan.
\newblock Head-and-neck squamous cell carcinoma patients with {CT} taken during
  pre-treatment, mid-treatment, and post-treatment ({HNSCC-3DCT-RT}).
\newblock \url{https://doi.org/10.7937/K9/TCIA.2018.13upr2xf}.

\bibitem[NW06]{NocedalWright:2006}
J.~Nocedal and S.~J. Wright.
\newblock {\em Numerical Optimization}.
\newblock Springer-Verlag, 2006.

\bibitem[Pag12]{Paganetti:2012}
H.~Paganetti.
\newblock Range uncertainties in proton therapy and the role of {M}onte {C}arlo
  simulations.
\newblock {\em Physics in Medicine and Biology}, 57(11):R99--R117, 2012.

\bibitem[PB14]{ParikhBoyd:2014}
N.~Parikh and S.~Boyd.
\newblock Block splitting for distributed optimization.
\newblock {\em Mathematical Programming Computation}, 6(1):77--102, 2014.

\bibitem[QvLLS16]{QiaoLew:2016}
Y.~Qiao, B.~van Lew, B.~P.~F. Lelieveldt, and M.~Staring.
\newblock Fast automatic step size estimation for gradient descent optimization
  of image registration.
\newblock {\em {IEEE} Transactions on Medical Imaging}, 35(2):391--403,
  February 2016.

\bibitem[Ray97]{Raydan:1997}
M.~Raydan.
\newblock The {B}arzilai and {B}orwein gradient method for the large scale
  unconstrained minimization problem.
\newblock {\em {SIAM} Journal of Optimization}, 7(1):26--33, 1997.

\bibitem[RN04]{RabbatNowak:2004}
M.~Rabbat and R.~Nowak.
\newblock Distributed optimization in sensor networks.
\newblock In {\em Proceedings of the 3rd International Symposium on Information
  Processing in Sensor Networks}, pages 20--27, April 2004.

\bibitem[RT16]{RamdasTibshirani:2016}
A.~Ramdas and R.~J. Tibshirani.
\newblock Fast and flexible {ADMM} algorithms for trend filtering.
\newblock {\em Journal of Computational and Graphical Statistics},
  25(3):839--858, 2016.

\bibitem[RTB04]{RaffardBoyd:2004}
R.~Raffard, C.~Tomlin, and S.~Boyd.
\newblock Distributed optimization for cooperative agents: Application to
  formation flight.
\newblock In {\em Proceedings of the 43rd {IEEE} Conference on Decision and
  Control ({CDC})}, pages 2453--2459, December 2004.

\bibitem[Say14]{Sayed:2014}
A.~H. Sayed.
\newblock Adaptation, learning, and optimization over networks.
\newblock {\em Foundations and Trends in Machine Learning}, 7(4--5):311--801,
  2014.

\bibitem[SLY15]{SunYe:2015}
R.~Sun, Z.-Q. Luo, and Y.~Ye.
\newblock On the expected convergence of randomly permuted {ADMM}, 2015.
\newblock arXiv:1503.06387.

\bibitem[SXSA14]{SawatzkyXu:2014}
A.~Sawatzky, Q.~Xu, C.~O. Schirra, and M.~A. Anastasio.
\newblock Proximal {ADMM} for multi-channel image reconstruction in spectral
  x-ray {CT}.
\newblock {\em {IEEE} Transactions on Medical Imaging}, 33(8):1657--1668, 2014.

\bibitem[TBD{\etalchar{+}}18]{TaastiRichter:2018}
V.~T. Taasti, C.~B{\"a}umer, C.~V. Dahlgren, A.~J. Deisher, M.~Ellerbrock,
  J.~Free, J.~Gora, A.~Kozera, A.~J. Lomax, L.~De Marzi, S.~Molinelli, B.-K.~K.
  Teo, P.~Wohlfahrt, J.~B.~B. Petersen, L.~P. Muren, D.~C. Hansen, and
  C.~Richter.
\newblock Inter-centre variability of {CT}-based stopping-power prediction in
  particle therapy: Survey-based evaluation.
\newblock {\em Physics and Imaging in Radiation Oncology}, 6(1):25--30, 2018.

\bibitem[THDZ20]{Taasti:Hierarchical}
V.~T. Taasti, L.~Hong, J.~O. Deasy, and M.~Zarepisheh.
\newblock Automated proton treatment planning with robust optimization using
  constrained hierarchical optimization.
\newblock {\em Medical Physics}, 47(7):2779--2790, 2020.

\bibitem[TMDQ16]{TanMa:2016}
C.~Tan, S.~Ma, Y.-H. Dai, and Y.~Qian.
\newblock Barzilai-{B}orwein step size for stochastic gradient descent.
\newblock In {\em Proceedings of the 30th Conference on Neural Information
  Processing Systems ({NIPS})}, volume~29, 2016.

\bibitem[UAB{\etalchar{+}}18]{Unkelbach:2018}
J.~Unkelbach, M.~Alber, M.~Bangert, R.~Bokrantz, T.~C.~Y. Chan, J.~O. Deasy,
  A.~Fredriksson, B.~L. Gorissen, M.~van Herk, W.~Liu, H.~Mahmoudzadeh,
  O.~Nohadani, J.~V. Siebers, M.~Witte, and H.~Xu.
\newblock Robust radiotherapy planning.
\newblock {\em Physics in Medicine and Biology}, 63(22):22TR02, 2018.

\bibitem[WCW{\etalchar{+}}17]{MatRad}
H.-P. Wieser, E.~Cisternas, N.~Wahl, S.~Ulrich, A.~Stadler, H.~Mescher, L.-R.
  M{\"u}ller, T.~Klinge, H.~Gabrys, L.~Burigo, A.~Mairani, S.~Ecker,
  B.~Ackermann, M.~Ellerbrock, K.~Parodi, O.~J{\"a}kel, and M.~Bangert.
\newblock Development of the open-source dose calculation and optimization
  toolkit {matRad}.
\newblock {\em Medical Physics}, 44(6):2556--2568, 2017.

\bibitem[WWG{\etalchar{+}}18]{MatRad:Int}
H.P. Wieser, N.~Wahl, H.S. Gabry{\'s}, L.R. M{\"u}ller, G.~Pezzano, J.~Winter,
  S.~Ulrich, L.~Burigo, O.~J{\"a}kel, and M.~Bangert.
\newblock {MatRad} – an open-source treatment planning toolkit for
  educational purposes.
\newblock {\em Medical Physics International Journal}, 6(1):119--127, 2018.

\bibitem[WWW17]{WangWu:2017}
Y.~Wang, S.~Wang, and L.~Wu.
\newblock Distributed optimization approaches for emerging power systems
  operation: A review.
\newblock {\em Electric Power Systems Research}, 144(1):127--135, 2017.

\bibitem[XFY{\etalchar{+}}17]{XuGoldstein:2017}
Z.~Xu, M.~Figueiredo, X.~Yuan, C.~Studer, and T.~Goldstein.
\newblock Adaptive relaxed {ADMM}: Convergence theory and practical
  implementation.
\newblock In {\em {IEEE} Conference on Computer Vision and Pattern Recognition
  ({CVPR})}, pages 7389--7398, July 2017.

\bibitem[XLLY17]{XuYang:2017}
Y.~Xu, M.~Liu, Q.~Lin, and T.~Yang.
\newblock {ADMM} without a fixed penalty parameter: Faster convergence with new
  adaptive penalization.
\newblock In {\em Advances in Neural Information Processing Systems ({NIPS})},
  volume~30, 2017.

\bibitem[YTL{\etalchar{+}}14]{YaoTempleton:2014}
R.~Yao, A.~Templeton, Y.~Liao, J.~Turian, K.~Kiel, and J.~Chu.
\newblock Optimization for high-dose-rate brachytherapy of cervical cancer with
  adaptive simulated annealing and gradient descent.
\newblock {\em Brachytherapy}, 13(4):352--360, 2014.

\bibitem[YYW{\etalchar{+}}19]{YangJohansson:2019}
T.~Yang, X.~Yi, J.~Wu, Y.~Yuan, D.~Wu, Z.~Meng, Y.~Hong, H.~Wang, Z.~Lin, and
  K.~H. Johansson.
\newblock A survey of distributed optimization.
\newblock {\em Annual Reviews in Control}, 47:278--305, 2019.

\bibitem[ZLYX15]{ZarepishehXing:2015}
M.~Zarepisheh, R.~Li, Y.~Ye, and L.~Xing.
\newblock Simultaneous beam sampling and aperture shape optimization for
  {SPORT}.
\newblock {\em Medical Physics}, 42(2):1012--1022, 2015.

\bibitem[ZXY18]{ZarepishehXing:2018}
M.~Zarepisheh, L.~Xing, and Y.~Ye.
\newblock A computation study on an integrated alternating direction method of
  multipliers for large scale optimization.
\newblock {\em Optimization Letters}, 12(1):3--15, 2018.

\end{thebibliography}

\pagebreak

\appendix

\section{ADMM Subproblems}
\label{app:admm_sub}

We are interested in solving Problem (\ref{prob:admm_sub}):
\[
\mbox{minimize}~ \sum_{i=1}^m w_i\|a_{s,i}^Tx_s - p_i\|_2^2 + \frac{\rho}{2}\|x_s - z^{(k)} + y_s^{(k)}/\rho\|_2^2
\]
with respect to $x_s \in \mathbf{R}^n$, where $z^{(k)} \in \mathbf{R}_+^n$, $y_s^{(k)} \in \mathbf{R}^n$, and $\rho > 0$ is a parameter. This is an unconstrained least-squares problem. Expand the objective function out to get
\begin{align*}
	&\sum_{i=1}^m w_i\|a_{s,i}^Tx_s - p_i\|_2^2 + \frac{\rho}{2}\|x_s - z^{(k)} + y_s^{(k)}/\rho\|_2^2 \\
	=& \|\tilde A_sx_s - \tilde p\|_2^2 + \left\|\sqrt{\frac{\rho}{2}}x_s - \sqrt{\frac{\rho}{2}}(z^{(k)} - y_s^{(k)}/\rho)\right\|_2^2 \\
	=& \left\|\left(\begin{array}{c}
		\tilde A_s \\
		\sqrt{\frac{\rho}{2}} I
	\end{array}\right)x_s - 
	\left(\begin{array}{c} 
		\tilde p \\
		\sqrt{\frac{\rho}{2}}(z^{(k)} - y_s^{(k)}/\rho)
	\end{array}\right)\right\|_2^2,
\end{align*}
where we define
\[
\tilde A_s := \textbf{diag}(\sqrt{w})A_s, \quad \tilde p := \textbf{diag}(\sqrt{w})p.
\]
Here $\textbf{diag}(\sqrt{w})$ is a diagonal matrix with $(\sqrt{w_1},\ldots,\sqrt{w_m})$ on the diagonal. 

By the normal equations, a solution $x_s^{(k+1)}$ of Problem (\ref{prob:admm_sub}) must satisfy
\begin{equation}
	\label{eq:admm_sub_normal}
	\underbrace{\left(\tilde A_s^T \tilde A_s + \frac{\rho}{2} I\right)}_{B_s}x_s^{(k+1)} = \underbrace{\tilde A_s^T \tilde p + \frac{\rho}{2}\left(z^{(k)} - y_s^{(k)}/\rho\right)}_{c_s^{(k)}}.
\end{equation}
Since $B_s$ is positive definite (because $\rho > 0$), such a solution always exists. One way to find $x_s^{(k+1)}$ is to form the Cholesky decomposition of $B_s$ and apply backward substitution to the triangular matrices. More precisely, we find an upper triangular matrix $U \in \mathbf{R}^{n \times n}$ that satisfies $B_s = UU^T$. Then, we solve the system of equations $Uv = c_s^{(k)}$ for $v$ and $U^Tx_s^{(k+1)} = v$ for $x_s^{(k+1)}$. These two solves can be done very quickly via backward substitution. Notice that $U$ need only be computed once at the start of ADMM, as $B_s$ remains the same across iterations, so subsequent solves of Problem (\ref{prob:admm_sub}) simply require us to update $c_s^{(k)}$ and solve the two triangular systems of equations.
	
\end{document}